\documentclass[twocolumn]{aastex631}

\usepackage{subfigure}
\usepackage{soul}
\usepackage{etoolbox}
\robustify{\cite}
\robustify{\citep}
\usepackage[normalem]{ulem}

\begin{document}

\title{Chemically primitive dwarf accretion reignites the inner disk assembly of Malin 1}

\author[0009-0008-2970-9845]{Manish Kataria}\thanks{E-mail: manish@iucaa.in}
\author[0000-0002-8768-9298]{Kanak Saha}\thanks{E-mail: kanak@iucaa.in}
\affiliation{Inter-University Centre for Astronomy and Astrophysics, Pune 411007, India.}

\begin{abstract}
We present a detailed kinematic and stellar population analysis of the inner disk of Malin 1, a giant low surface brightness (GLSB) galaxy with a prominent SB0-type central morphology. AstroSat far-UV imaging reveals clumpy emission features indicating recent star formation. Using MUSE integral field spectroscopy, we identify four star-forming complexes (SFCs) within the inner 10 kpc, each associated with localized ionized gas emission in distinct H$\alpha$ velocity channels. Two of the SFCs, including a far-UV clump, appear on the blue-shifted side ($V_{H\alpha}=-230\ \mathrm{kms^{-1}}$), while the other two are redshifted. The far-UV clump shows a strong velocity offset ($\sim150\ \mathrm{kms^{-1}}$) and high gas dispersion ($\sim250\ \mathrm{kms^{-1}}$), indicating that it is kinematically decoupled from the rotating disk. The spatial and velocity isolation of these features in the channel map confirms they do not follow regular disk rotation. 

The far-UV clump hosts young (250–500 Myr), extremely metal-poor ([M/H]$\simeq$ –1.69) and $\alpha$-enriched ($[\alpha/Fe] \sim 0.5$) stellar populations, sharply contrasting with the surrounding super-solar gas-phase metallicity. The young stellar populations in each SFC are chemically distinct (similar to the far-UV clump) from the enriched central ISM, indicating rapid, local star formation from primitive gas before efficient mixing with the enriched ISM. Their spatial and velocity segregation, age synchronicity, and chemical homogeneity suggest an origin of gas delivered by a disrupted, gas-rich dwarf on a high-inclination (off-plane) orbit. These results suggest that the central HSB within $\rm \sim 9^{\prime\prime} (14\ kpc)$ radius component of Malin 1 has grown through discrete, externally driven accretion, contributing to its complex, hybrid disk morphology.
\end{abstract}


\section{Introduction}\label{sec:intro}

Giant Low Surface Brightness (GLSB) galaxies represent some of the most extreme cases of galaxy formation and evolution, characterized by stellar disks extending up to several hundred kiloparsecs and central surface brightness fainter than $\mu_B \approx 23\ \mathrm{mag\ arcsec^{-2}}$ \citep{Impey&Bothun1997}. Often classified as spirals due to their prominent arms \citep{Hubble1926}, GLSBs differ from typical high surface brightness (HSB) spirals in several key ways: their spiral arms extend to much larger radii, they exhibit unusually low star formation rates despite hosting large H{\sc i} reservoirs ($M_{\rm HI} \gtrsim 10^{10}\ M_\odot$; \citep{Matthews2001,O'Neil2004}), and they often reside in isolated, low-density environments. These galaxies serve as natural laboratories for studying alternative evolutionary pathways, shaped by inefficient star formation, prolonged gas accretion, and minimal dynamical disturbance \citep{Das2013, Martin2019, Young2015}.

Despite these insights, the origin and long-term stability of such diffuse, massive disks remain poorly understood. Key questions include how these systems prevent fragmentation despite their low surface densities \citep{Dalcanton1997, Mapelli2008}, and how some develop a hybrid morphology, where a HSB central component coexists with a giant LSB disk, as seen in Malin 1 \citep{Pickering1999, Saha2021, SarkarSaha2025}. Their low star formation efficiency despite ample gas content \citep{Bothunetal1987, Hoffman1992} and their structural diversity from undisturbed disks to signs of past interactions \citep{Noguchi2001, Penarrubia2006} complicate efforts to define a unified formation scenario. Their rarity in the local universe, coupled with their strong dark matter dominance, suggests that standard galaxy formation models may be inadequate \citep{Kasparova2014}, pointing instead to alternative formation scenarios involving high spin halos, late infall, or suppressed star formation over cosmic time.

Among GLSBs, Malin 1 stands out as the most iconic and extreme example, and was the first such galaxy to be identified \citep{Bothunetal1987}. Deep optical imaging with the Canada-France-Hawaii Telescope (CFHT) revealed a giant, diffuse stellar disk extending nearly 160 kpc, about six times the size of the Milky Way stellar disk \citep{Galaz2015}. Along with an extended outer disk, Malin 1 hosts a compact, HSB inner region spanning 14 kpc in radius, structurally consistent with a barred lenticular (SB0) galaxy, featuring a pseudobulge, an exponential disk, and a prominent $\sim$6 kpc bar \citep{Barth2007, Saha2021}. In other words, the inner HSB disk resembles the morphology of a quiescent early-type system rather than a typical star-forming spiral. This hybrid structure (a central HSB and outer LSB disk) already challenges conventional disk formation scenarios and has motivated efforts to probe its central regions in greater detail.

Although the inner region of Malin 1 has been classified as a barred lenticular (SB0) galaxy, redder g-i color indicating an older underlying stellar population, multiple evidence suggest that it is not a passively evolving system. Ultraviolet imaging with GALEX first revealed FUV and NUV emission from the central region, indicating ongoing or recent star formation \citep{Boselli2011}. Follow-up high-resolution UV imaging with AstroSat/UVIT in the F154W (FUV) and N263M (NUV) bands uncovered compact, clumpy star-forming regions distributed asymmetrically across the central region and, in particular, an FUV-bright clump (not present in N263M imaging) on the south-east (SE) side, close to the bar major axis \citep{Saha2021}. Besides, the presence of a strong $\rm H\alpha$ emission in the SDSS 3$^{\prime\prime}$ aperture spectrum points to a recent burst of star formation in the same region. Not just the central region, the outer LSB disk displays tightly wound spiral arms with star-forming clumps as revealed by CFHT deep imaging and UVIT. These regions, including the prominent central FUV clump, have been examined in detail using MUSE integral field spectroscopy \citep{Junais2024}. A careful spectral analysis by \cite{Johnston2024} revealed an additional compact source in the H$\rm \alpha$ narrow band image, located north-west of the central clump along the bar’s major axis. They interpret this as an off-center compact source, possibly falling into the center, and suggest an external origin as one of the possibilities.

However, directly confirming the external origin of these star-forming clumps remains challenging, as their spectral signatures are often overwhelmed by the light from the dominant older stellar populations. Nevertheless, several companion galaxies, most notably Malin 1B and SDSS J123708.91+142253.2, have been spectroscopically confirmed to be interacting with Malin 1, despite its location in an extremely under-dense environment \citep{Reshetnikovetal2010, Galaz2015}. Interestingly, similar scenarios have emerged in cosmological simulations such as IllustrisTNG, where Malin 1 analogues are found to undergo minor mergers with gas-rich dwarfs, leading to renewed star formation activity and contributing to disk growth \citep{Zhu2018}. Motivated by these findings, we perform a detailed investigation of the kinematics and stellar populations in the central region of Malin 1, including the prominent star-forming clumps, using archival MUSE integral field spectroscopy. The remainder of this paper focuses on uncovering the origin of these clumps and their implications for the late-stage assembly of Malin 1.
 
This paper is formulated as follows. We describe the AstroSat and MUSE observations, of Malin 1 in Section \ref{sec:obs}. In Section \ref{sec:kinemtaics}, we outlined the various kinematic analyses carried out using MUSE data. Section \ref{sec:chemistry} comprises results obtained from stellar population analysis, and finally, in Section \ref{sec:conclusion}, we conclude our findings on the formation history of Malin 1.

Throughout this paper, we adopt the flat $\Lambda$CDM cosmology with $\rm H_{o}=70 Kms^{-1}Mpc^{-1},\ \Omega_{M} = 0.3 \ and \ \Omega_{vac} = 0.7$.

\begin{figure*}[ht!]
\includegraphics[width = 6.5in]{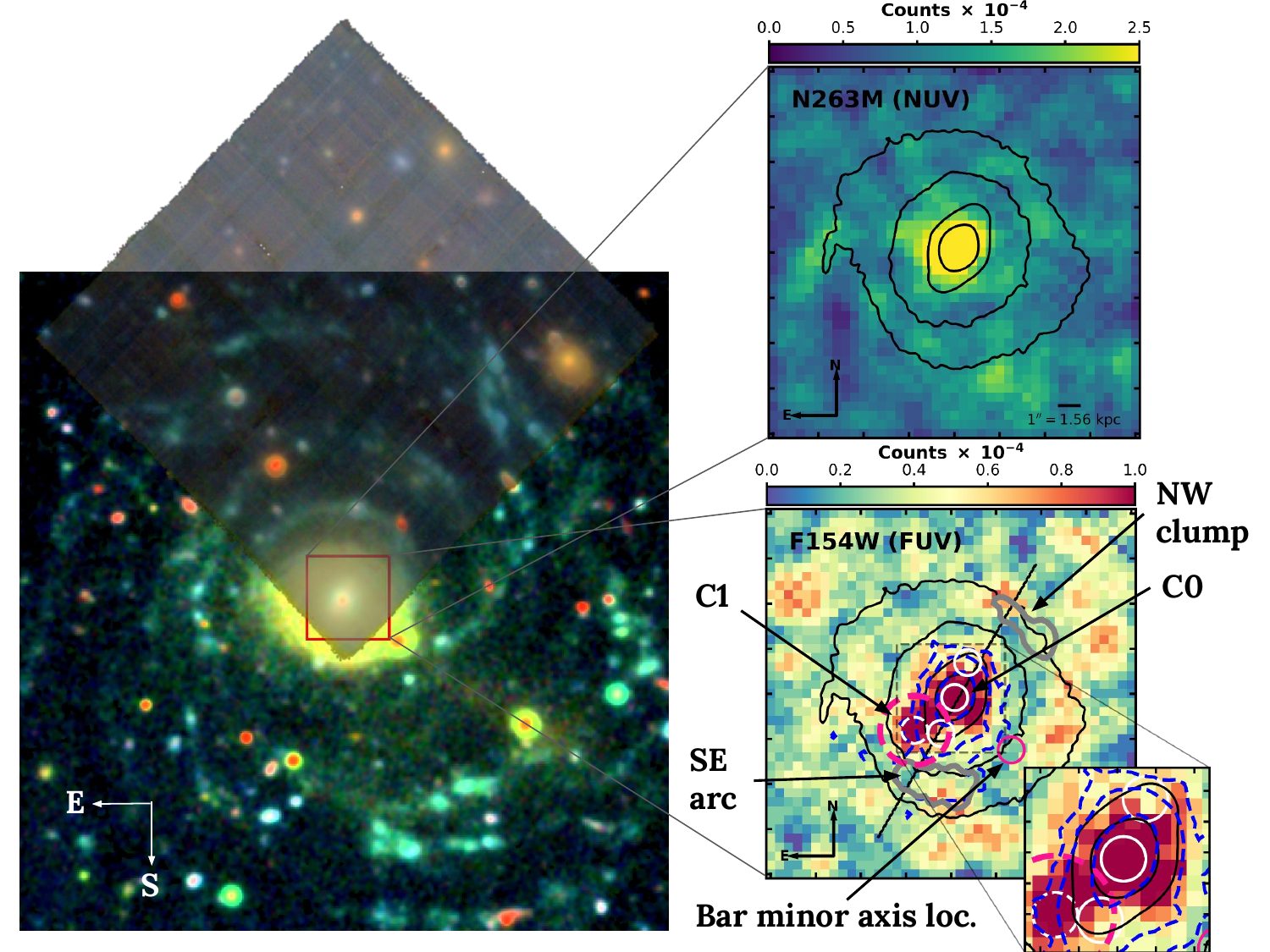}
\caption{\textbf{Left:} CFHT (i+g+u) RGB image of Malin 1 overlaid with MUSE RGB, showing the spiral structure. The red square marks the MUSE cutout used for analysis.\textbf{Right:} Astrosat/UVIT F154W (FUV) and N263M (NUV) smoothed images. FUV reveals the central source (C0) and an offset clump (C1), while only C0 is seen in NUV. Black contours: HST/WFPC2 F814W stellar emission. Gray contours: $\rm H\alpha$ channel map contours of clumps detected at different velocities. Blue dashed contours (1, 3, and 15$\sigma$): $\rm H_{\alpha}$ emission at $\rm V_{H\alpha}=-230\ \rm kms^{-1}$. White circles ($r=0.6^{\prime\prime}$): MUSE spectral extraction regions. Pink dashed circle ($r=1.6^{\prime\prime}$): UVIT/F154W PSF FWHM; solid red circle: bar minor-axis extraction region. Black line: bar orientation from HST. Zoom-in highlights the clumpy FUV extended region, and asymmetric $\rm H\alpha$ emission relative to the bar.
\label{fig:malin1_rgb_fuv}}
\end{figure*}

\begin{figure*}[ht!]
\centering
\includegraphics[width = 6.5in]{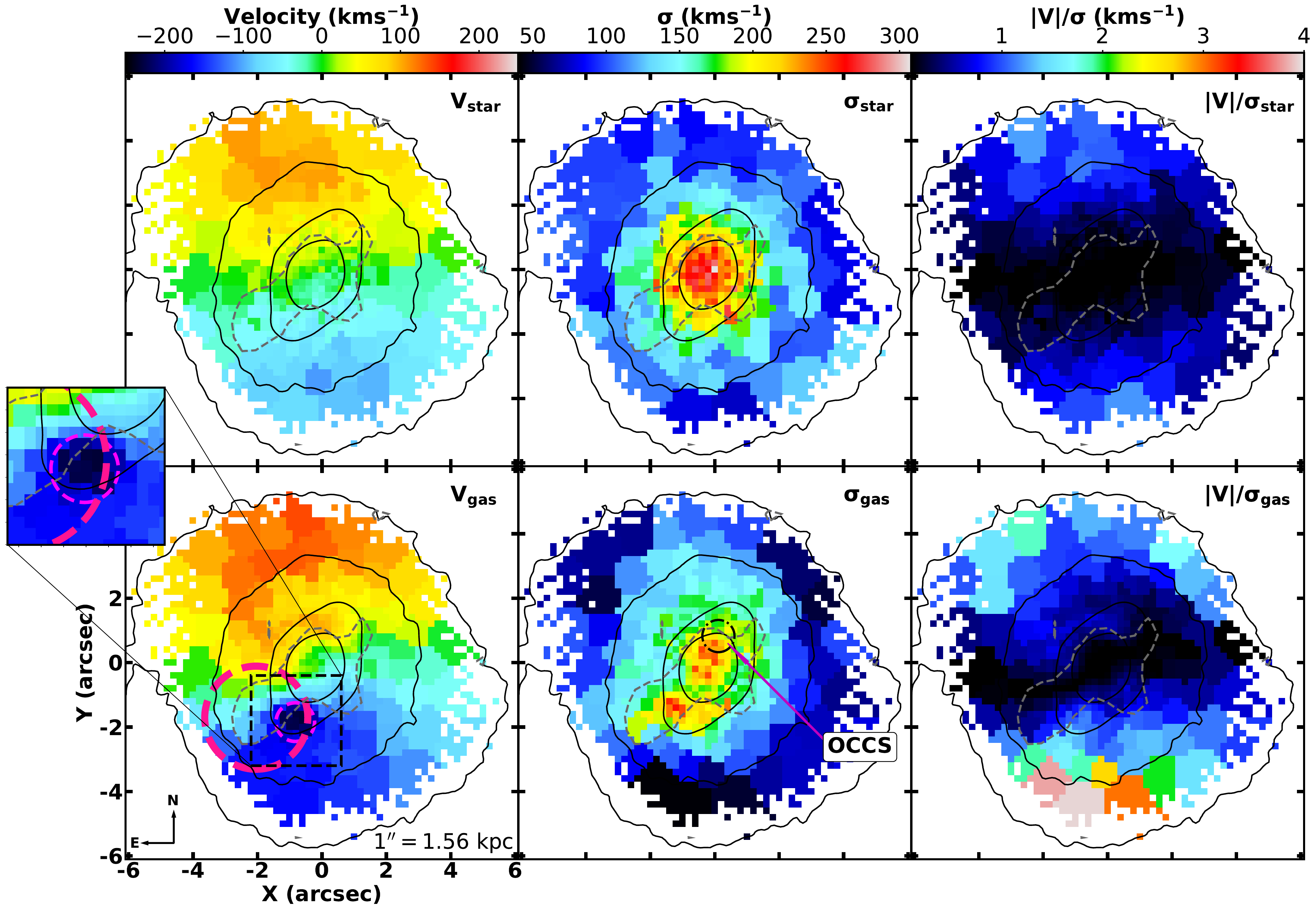}
\caption{Kinematic maps of the central HSB region, from MUSE IFU data, of Malin 1 (red square in the left panel of Figure \ref{fig:malin1_rgb_fuv}). The solid black and dashed gray contours are HST/WFPC2 F814W and Astrosat/UVIT F154W ($2\sigma$), respectively. \textbf{Top panel: }stellar kinematics maps: left to right panel shows the velocity (V), dispersion ($\sigma$), and $|V|/\sigma$ maps. \textbf{Bottom panel: }ionized gas kinematics: \textit{left:} gas velocity maps with dashed magenta circle ($r = 0.6^{\prime\prime}$) show the velocity enhanced region, the deepink dashed circle is centered at the location of C1 (see Figure~\ref{fig:malin1_rgb_fuv}) with $r = 1.6^{\prime\prime}$ corresponds to the FWHM of the FUV. The velocity-enhanced region is highlighted in the zoom-in. \textit{Middle:} Shows the velocity dispersion with the black dot-dashed circle showing the off-center compact source (OCCS, \citep{Johnston2024}) and two distinct regions of high dispersion, \textit{right:} shows the $\rm |V|/\sigma$ with a high $\rm |V|/\sigma \sim 4$ region in corresponding to SE arc (see Figure~\ref{fig:malin1_ha_channel}).
\label{fig:malin1_velocity_maps}}
\end{figure*}

\begin{figure*}[ht!]
\centering
\includegraphics[width = 6.5in]{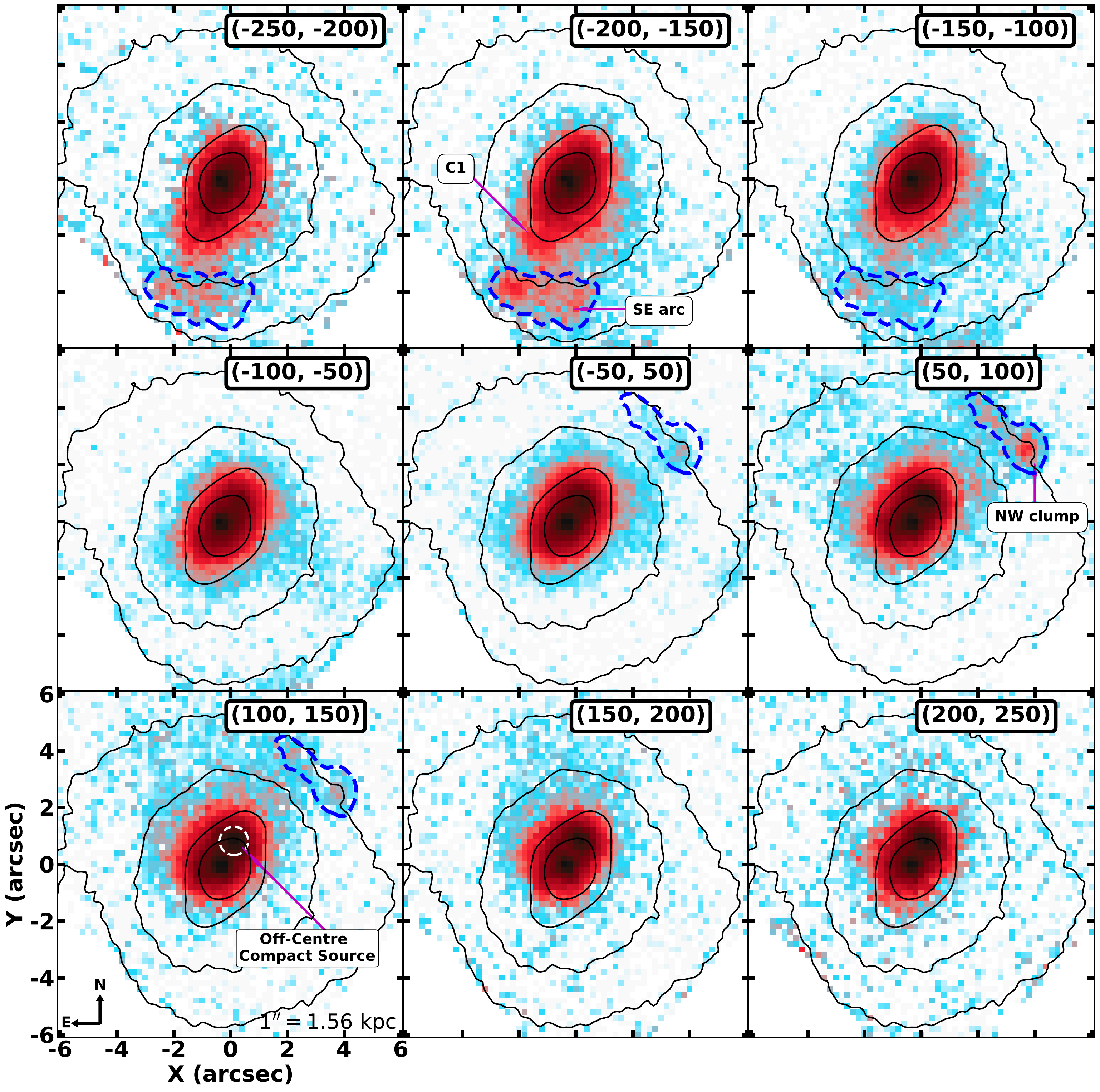}
\caption{$\rm H\alpha$ channel maps from -250 to 250 $\rm kms^{-1}$ with $\rm 50\ kms^{-1}$ bins consisting of a single MUSE wavelength slice (except the central -50 to 50 $\rm kms^{-1}$) showing different features. The top middle panel (-200, -150) shows the clump C1 and SE arc clearly. The middle right panel (50, 100) shows an extended feature called the NW clump. The bottom panel shows the OCCS clump with the highest contrast, although it is visible in most of the channel maps (-50, 250).
\label{fig:malin1_ha_channel}}
\end{figure*}

\section{AstroSat and MUSE observations}
\label{sec:obs}
We make use of archival observations from AstroSat/UVIT and VLT/MUSE in this study. Malin 1 was observed with UVIT in the FUV (F154W) and NUV (N263M) bands, providing high-resolution ultraviolet imaging of both the central and extended regions. Details of the UVIT observing strategy, exposure times, and data reduction are described in \citet{Saha2021}. Figure~\ref{fig:malin1_rgb_fuv} shows the central region of the galaxy, overlaid with the MUSE field of view (spatial resolution $\sim0.6^{\prime\prime}$) on the CFHT RGB image. The MUSE observations, originally presented in \citet{Junais2024} and \citet{Johnston2024}, are fully described in those works. However, in order to present a self-contained analysis of the MUSE observations, we re-run the pipeline with slight modifications that helped reduce the noise in the redder side of the spectral window (see appendix for details). MUSE provides the observed spectral coverage within the optical regime (4800-9300 \text{\AA}) over a field of view of $1^{\prime}\times1^{\prime}$ with a spatial sampling of $0.2^{\prime\prime}\ pix^{-1}$ in Wide Field Mode (WFM).

The UVIT N263M image shows a smooth, centrally concentrated emission, while the F154W image reveals several clumpy structures indicative of recent star formation. Among these, a particularly prominent FUV-bright clump appears at the tip of the bar toward the southeast (SE) (see Figure~\ref{fig:malin1_rgb_fuv}), marked as C1. Whether these central FUV clumps formed in situ or have migrated from elsewhere remains unclear. This study aims to investigate the origin of these clumps and the possible ongoing assembly of Malin 1’s central region, which otherwise appears dominated by an old stellar population. Building upon the UVIT and MUSE datasets, we conduct a detailed kinematic and stellar population analysis of the ionized gas and young stellar components in the central disk.

\section{Stellar and Ionized gas kinematics}\label{sec:kinemtaics}

The stellar velocity map (Figure~\ref{fig:malin1_velocity_maps}) shows a well-defined, rotating disk in the central HSB region of Malin 1, with ${V_{min}, V_{max} \approx -106,\ 109\ \mathrm{km\ s^{-1}}}$. The velocity dispersion peaks at the bulge ($\sigma_* \approx 307\ \mathrm{km\ s^{-1}}$) and declines smoothly to $\sim 75\ \mathrm{km\ s^{-1}}$ toward the outskirts typical of massive disk galaxies \citep{Oh2020}. The ${V/\sigma}_{LOS}$ map (Figure~\ref{fig:malin1_velocity_maps}, top right) indicates a dynamically hot, pressure-supported bulge, with $V/\sigma < 1.0$ within the inner region and never exceeding 1.2 across the HSB disk. This is significantly lower than typical S0s observed using planetary nebulae kinematics \citep{Cortesietal2013}, but consistent with field S0 galaxies observed with MUSE \citep{Coccatoetal2020} and simulations of minor-merger-driven S0 formation \citep{Bournaudetal2005}. The presence of ongoing minor interactions, including with Malin 1B and another companion at $\sim$350 kpc \citep{Reshetnikovetal2010}, further supports a merger-driven origin for the S0-like inner disk.
The ionized gas velocity map (Figure~\ref{fig:malin1_velocity_maps}, bottom left) largely follows the stellar rotation, but exhibits a localized velocity excess ($\rm \sim150\ km\ s^{-1}$) near the SE C1 clump. This feature is co-located with FUV emission in the F154W image and lies within the UVIT PSF (Figure~\ref{fig:malin1_rgb_fuv}). To quantify the anomaly, we extracted spectra from a $1.2^{\prime\prime}$ aperture centered on C1 and a symmetric region on the opposite side of the bar. The clump exhibits a LOS velocity of $\sim 230\ \mathrm{km\ s^{-1}}$ (approaching), while the comparison region (opposite end of the bar) shows only $\sim 79\ \mathrm{km\ s^{-1}}$ (receding), reinforcing the presence of a kinematically distinct component (Figure~\ref{fig:malin1_extracted_spectra}).

Interestingly, this anomaly is absent in the stellar velocity map, likely due to the young stars in C1 contributing little to the total continuum, thus not affecting the stellar absorption features \citep{Rosado-Belza2020}. In contrast, ionized gas emission powered by recent star formation stands out and traces the kinematics of this younger component.

The ionized gas velocity dispersion map (Figure~\ref{fig:malin1_velocity_maps}, bottom middle) shows two peaks: one at the center and another in the SE region, coinciding with C1. Elsewhere, dispersion declines smoothly, mimicking the stellar profile. While the stellar $V/\sigma$ remains below 1 across the disk, the gas $V/\sigma$ exceeds 3 in localized regions (Figure~\ref{fig:malin1_velocity_maps}, bottom right) in the SE arc (highlighted in the top panels of Figure~\ref{fig:malin1_ha_channel} $\rm H\alpha$ channel maps). This one-to-one mapping between velocity anomaly and enhanced gas dispersion with no corresponding feature in the stars points to a dynamically decoupled, likely external origin for the C1 clump. In an ideal case, gas inflow driven purely by a bar potential (i.e., a bi-symmetric potential) should follow a symmetric pattern around the bar. The gas density is usually low within most of the bar region, with enhanced densities forming along the bar edges and in nuclear spiral-like structures \citep{Athanassoula1992}. Therefore, a bar-driven inflow would lead to density enhancements on both sides of the bar rather than a global asymmetry. However, in Malin 1, we observe asymmetric FUV emission, with the C1 clump appearing only on the south-east side of the bar. This indicates that the observed inflow is unlikely to be caused solely by the bar potential.

\subsection{$H\alpha$ velocity channel maps from MUSE data} 
\label{subsec:channel_maps}

Despite Malin 1's large H{\sc i} reservoir ($\rm M_{HI} \approx 7\times10^{10}\ M_\odot$; \citealt{Pickeringetal1997}) and multiple FUV-bright clumps indicating recent star formation, the MUSE cube collapsed around H$\alpha$ shows a smooth, S0-like structure. This is surprising, especially since MUSE has higher spatial resolution ($\sim$0.6$^{\prime\prime}$) than UVIT. This discrepancy of not having a similar clumpy structure in the collapsed cube around $\rm H\alpha$ arises from the strong stellar continuum of the dominant old population, which overshines the relatively weak H$\alpha$ emission. To overcome this, we examine velocity-channel maps of H$\alpha$, where kinematically distinct features can emerge from beneath the continuum. We model the stellar continuum in each spaxel across the central $12^{\prime\prime}\times12^{\prime\prime}$ region using pPXF \citep{Cappellari2017} and subtract it from the original spectra to produce a continuum-subtracted cube. pPXF is a full-spectrum fitting code to derive the stellar kinematics and population properties from the galaxy spectra. The pPXF algorithm fits an observed spectrum with a combination of template spectra convolved with a line-of-sight velocity distribution (LOSVD), while penalizing unphysical solutions through regularization. This technique enables accurate measurements of stellar velocity, velocity dispersion, and stellar population mix, even in spectra with moderate signal-to-noise ratios. From this emission line only cube, we constructed $\rm H\alpha$ channel maps at velocities from $V = -250\ \text{to}\ +250 \ \mathrm{km\ s^{-1}}$ in steps of $50 \ \mathrm{km\ s^{-1}}$ and they are arranged from blue-shifted side to the red-shifted side of the galaxy (see Figure~\ref{fig:malin1_ha_channel}). The $\rm H\alpha$ channel maps reveal four distinct features appearing exclusively in particular velocity bins, and these are described below:

\noindent \textbf{C1 clump:}\\ 
This appears most prominently in the $V = -200\ \text{to}\ -150\ \mathrm{km\ s^{-1}}$ channel (specifically at $V = -230\ \mathrm{km\ s^{-1}}$) at distance of $\sim 3\ \text{kpc}$ from the centre. This feature spatially coincides with the location of the velocity anomaly and FUV clump (see F154W map and ionized gas velocity map), suggesting that the same star-forming region is traced both in the ionized gas and the young stellar continuum, with some positional offset that may arise from the influence of the galaxy’s differential rotation, plus differences in the physical timescales probed by $\rm H\alpha$ (very recent, $\lesssim 10$ Myr) and FUV (up to $\sim 100$ Myr) emission. This feature is absent in the rest of the velocity channels (starting $V = -100\ \text{to}\ -50\ \mathrm{km\ s^{-1}}$) and hints towards a sign of an ongoing interaction in the central region. C1 clump exhibits a significant velocity offset ($\rm 150\ km\ s^{-1}$) from the surrounding disk rotation and coincides with a region of high H$\rm \alpha$ velocity dispersion ($\sim 250\ \mathrm{km\ s^{-1}}$), making this clumpy feature kinematically decoupled from the surrounding disk region.

\noindent \textbf{Sout-East Arc (SE arc):}\\
Like the C1 clump, SE arc is most prominently visible in the $V = -200\ \text{to}\ -150\ \mathrm{km\ s^{-1}}$ channel and located at a distance of $\sim 6\ \text{kpc}$ from the centre towards the South-East direction.  SE arc appears as nearly linear structure, approximately 4-5 kpc in projected length, oriented along the east–west direction. It appears in the adjacent velocity slices, with the brightest emission localized toward its eastern end. The linear, asymmetric brightness distribution suggests a stream of filament-like structure, possibly tracing ionized gas inflow. Like C1, this feature is absent in the rest of the velocity channels (starting $V = -100\ \text{to}\ -50\ \mathrm{km\ s^{-1}}$). Although the LOS velocity in the SE arc region is similar to the local disk rotation, it's velocity dispersion is only $\sigma_{gas}\simeq 40 - 50\ \mathrm{km\ s^{-1}}$, making the $V/\sigma_{gas} \sim 4$ and once again this feature stands out kinematically from the rest of the inner disk. 

\noindent \textbf{North-West clump (NW clump):}\\
This is a morphologically distinct and resembles a tadpole or cometary structure, located north-west of the nucleus at a projected distance of $\sim 7\ \text{kpc}$. This feature first appears in the  $V = -50\ \text{to}\ +50\ \mathrm{km\ s^{-1}}$ channel, becomes prominent with a clear head and tail extending toward the northeast at  $V = 50\ \text{to}\ 100\ \mathrm{km\ s^{-1}}$ slice and again goes fainter in  $V = 100\ \text{to}\ 150\ \mathrm{km\ s^{-1}}$ channel, with no trace in the rest of the channel maps. NW clump has velocity dispersion $\sim 50\ \mathrm{km\ s^{-1}}$ and LOS velocity consistent with surrounding disk material in its near vicinity.

\noindent \textbf{C0 clump:}\\
This is the central FUV clump (extracted within an aperture of size 0.6", see Figure~\ref{fig:malin1_rgb_fuv}) which contains part of the off-centre compact source (OCCS) as reported by \cite{Johnston2024}. The OCCS is visible mostly in the redshifted channels and appears most prominently (based on the $\rm H\alpha$ intensity) in the $V = 200\ \text{to}\ 250\ \mathrm{km\ s^{-1}}$ slice. As shown by \cite{Johnston2024}, the gas kinematics of OCCS is found to be consistent with that of the central compact source (CCS (C0)) and they are neighbours. However, the central gas velocity dispersion map suggests that OCCS has slightly higher velocity dispersion than C0. According to \cite{Johnston2024}, OCCS could be a star-forming clump that is falling onto the core of the galaxy or a clump of infalling gas either from outside or from the disc and triggered star formation at the current location.

\subsection{Gas-phase metallicity of the central disk} 
\label{subsec:gas_metallicity_extinction}

Motivated by the kinematic anomalies in Malin 1’s inner disk, we examined the spatial distribution of gas-phase metallicity using the N2 index \citep{PettiniPagel2004}. While earlier studies suggested subsolar metallicity, e.g., based on the non-detection of CO(1$\rightarrow$0) by \citet{ImpeyBothun1989}, or broad-band SED fits by \citet{Boissier2016}, these were limited in spatial resolution or spectral coverage and hence were not well constrained. More recently, \citet{Junaisetal2020}, using spectroscopic data, found that the central region has near-solar metallicity, about 0.15 dex higher than earlier estimates.

\begin{figure*}[ht!]
\centering
\includegraphics[clip, trim=0cm 1.5cm 0cm 0cm, width = 6in]{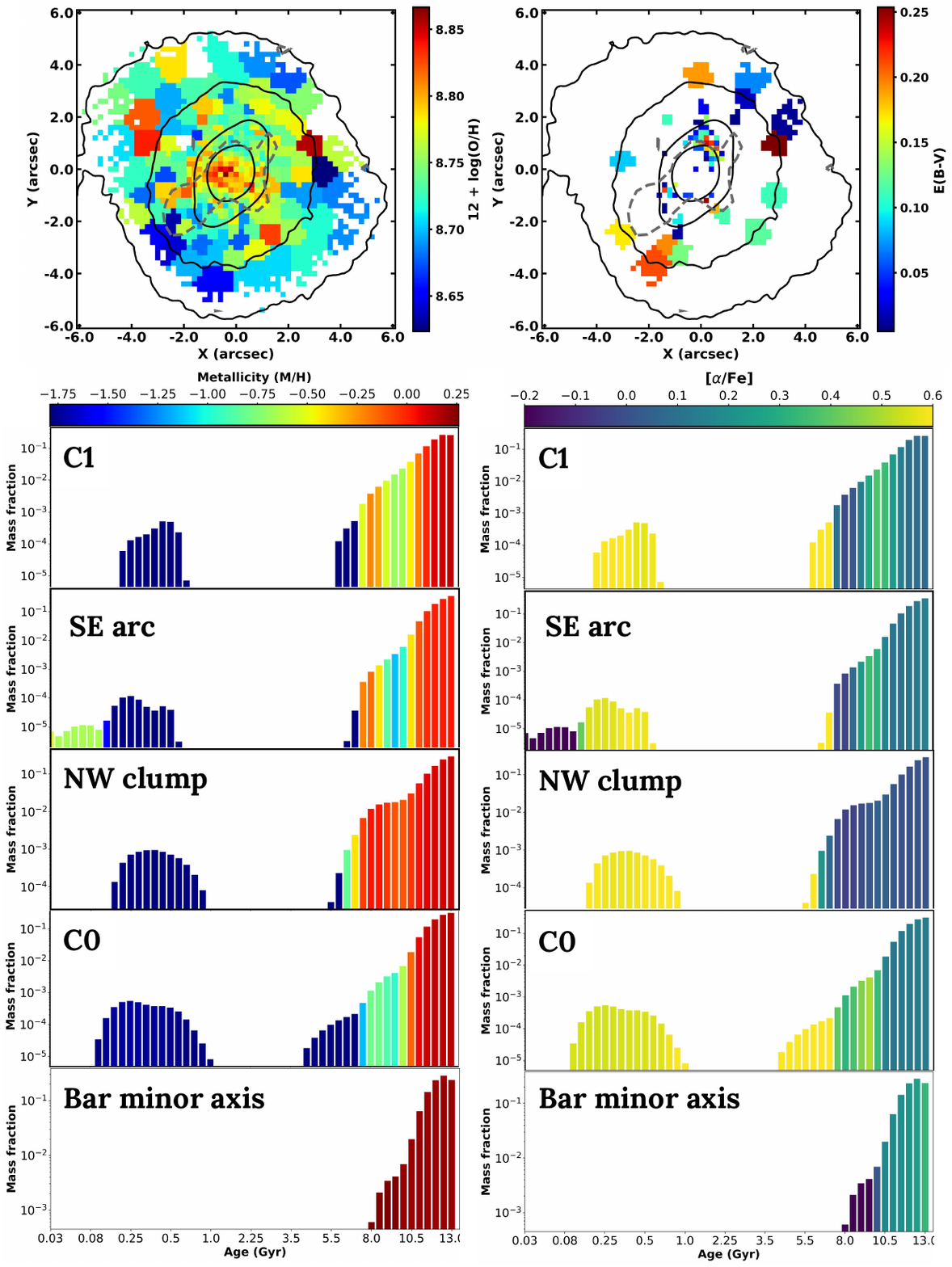}
\caption{\textbf{Top Panel: }Resolved gas metallicity (left) and the extinction, E(B-V) (right). With gas metallicity being $\gtrsim$ solar metallicity (12 + log(O/H) = 8.69) in most parts and supersolar, especially at the center. \textbf{Bottom Panel: } The left column shows the distribution of metallicity in the stellar population of different features identified in the central HSB disk, the right column shows the alpha-enrichment $[\alpha/Fe]$ for the same stellar population. The Bar minor axis sps is added for the comparison, where there is no FUV emission.
\label{fig:malin1_chem}}
\end{figure*}

We measure [N{\sc ii}] $\lambda$6583\AA\ and H$\alpha$ fluxes after subtracting the stellar continuum and absorption features from the underlying older stellar population using EMILES templates \citep{Vazdekis2016}. The cube was spatially binned to enhance S/N and ensure that the recovered emission-line fluxes were robust. The metallicity map (Figure~\ref{fig:malin1_chem}, (top left)) reveals a central peak of 12 + log(O/H) = 8.87 ($\sim$1.51 $Z_{\odot}$), decreasing slightly along the bar major axis to  12 + log(O/H) $\sim$ 8.75 ($\sim$1.15 $Z_{\odot}$). Nevertheless, the metallicity across the central region remains above solar, in agreement with the findings of \citet{Junaisetal2020} and confirm that Malin 1’s inner disk is chemically evolved, unlike most LSB galaxies \citep{McGaugh1994}. This supports a hybrid scenario: a chemically enriched HSB central region embedded within an extended, extremely diffuse LSB disk.

We derived a resolved gas-phase metallicity map using the O3N2 calibration \citep{Pettini2004} (see Figure~\ref{fig:malin1_o3n2_metallicity_map}, left). The resulting distribution is consistent with that obtained from the N2 method, with differences of at most $\sim0.2$ dex in $12 + \log(\mathrm{O/H})$. According to the emission-line diagnostic presented in \citet{Johnston2024}, most of the bins in the central region of Malin~1 fall between the \citet{Kewley2001} and \citet{Kauffmann2003} demarcation curves (i.e., within the composite region). Therefore, to obtain reliable gas-phase metallicities for the clumps analyzed in this work, we employed the HII-CHI-MISTRY code \citep{Perez-Montero2014, Perez-Montero2019}.

We measured the fluxes of the emission lines $\rm [OIII],\lambda5007$, $\rm [NII],\lambda6584$, and $\rm [SII],\lambda\lambda6717,6731$, each normalized to the $\rm H\beta$ flux. The uncertainties on the line fluxes were estimated using Monte Carlo (MC) realizations of the spectra, and all fluxes were corrected for extinction based on the observed $\rm H\alpha/H\beta$ ratio.

The code was executed under two ionization assumptions: (1) pure star formation, and (2) LINER-like ionization, following the prescriptions of \citet{Perez-Diaz2021}. In brief, for the star-forming case, we adopted the POPSTAR SED \citep{Molla2009}, which is a stellar population synthesis model with Chabrier IMF \citep{Chabrier2003} based on Padova isochrones having a wide range of 12  log[O/H] abundance from 6.9 to 9.1 with a spacing of 0.1 dex and the corresponding constraints for star-forming galaxies, while for the LINER case, we used an AGN double-component SED use to model the line emission with Big Blue Bump from accretion disc thermal emission and a power law having a spectral index, $\rm \alpha_{X} = -1$ with $\alpha_{\mathrm{OX}} = -0.8$ being the spectral index for xray to UV transition regime and a fraction of 2\% free electrons, consistent with low-ionization AGN conditions.

As shown in Figure~\ref{fig:malin1_o3n2_metallicity_map} (right), the metallicities derived under the star-forming assumption are consistent with those obtained from strong line methods, whereas the LINER-based models yield systematically lower values. Hence, the true metallicities of the clumps are likely to lie between these two limits, with an average value of $\rm \sim0.6~Z_{\odot}$ , which is the value when we calculate the mean of the average values of all four clumps from star-forming and LINER models. Furthermore, these regions appear extended or elongated in the $\rm H\alpha$ channel maps, indicating that they cannot arise solely from central emission. The same faint structures are also visible in the FUV image of the central region of Malin 1 (see Figure\ref{fig:malin1_rgb_fuv}), providing additional evidence for ongoing star formation.

We find no significant metallicity contrast at the C1 clump location, suggesting that any metal-poor, recently accreted gas is either chemically diluted or not dominant in emission. However, the clear presence of young stars in far-UV band implies recent star formation. To probe the origin of this population, we perform a detailed stellar population synthesis modeling (section~\ref{sec:chemistry}).

\section{Archaeology of Malin 1's central region} 
\label{sec:chemistry}

To investigate the stellar populations in the central region of Malin 1, we extracted combined spectra from selected apertures (Figure~\ref{fig:malin1_velocity_maps}) and fit them using the sMILES stellar population library \citep{Knowles2023} with the pPXF code \citep{Cappellari2017}. The sMILES models, with a wavelength range of 3540.6-7409.6 \text{\AA}, offer finer sampling in age and [$\alpha$/Fe] compared to EMILES \citep{Vazdekis2016}, with identical spectral resolution (2.5\AA). We adopt a Salpeter IMF and a regularization value of 0.05, following the approach in \citet{Pinna2019}.

Figure~\ref{fig:malin1_chem} reveals that the C1 clump underwent an early star formation episode about 6 Gyr ago, followed by a long quiescent phase and then a recent burst starting ~250 Myr ago, lasting another ~250 Myr. Beneath this recent burst lies an older stellar population (ages $>$6 Gyr) with mostly super-solar metallicities. Some stars in the 6–10 Gyr range are sub-solar, but stars older than 10.5 Gyr are again metal-rich ($Z > Z_{\odot}$).

In contrast, the young stars formed during the recent burst have extremely low metallicity ($[M/H] \sim -1.69$). This makes it unlikely that they formed from reprocessed gas in Malin 1's inner disk, which is chemically enriched (see Figure~\ref{fig:malin1_chem}, (top left)). Therefore, this episode cannot be a typical rejuvenation event driven by residual disk gas \citep{Barwayetal2020}.

The right panel of Figure~\ref{fig:malin1_chem} shows that the old, metal-rich population is $\alpha$-poor ($[\alpha/\mathrm{Fe}] < 0.1$), suggesting slow chemical evolution dominated by Type Ia supernovae. In contrast, the younger population is highly $\alpha$-enhanced, pointing to rapid enrichment by core-collapse supernovae. These chemical signatures low metallicity and high $\alpha$/Fe indicate that the gas fueling the recent star formation in C1 was both external and chemically unevolved. Combined with its kinematic decoupling, this strongly supports the idea that C1 is the remnant of an infalling disrupted dwarf.

Similar stellar population properties are seen in the SE arc and NW clump. The SE arc is also metal-poor and $\alpha$-enhanced like C1, but contains an even younger population (30–80 Myr) that is slightly more metal-rich ($Z \sim 0.18\ Z_\odot$) and less $\alpha$-enhanced, suggesting self-enrichment via mixing with the host ISM. The NW clump, like C1, is kinematically distinct and chemically primitive.

Interestingly, the C0 region (center of Malin 1) shows a similar pattern: metal-poor, $\alpha$-enhanced stars formed between 250 Myr and 1 Gyr ago, and an older metal-poor population formed $\sim 4 – 7$ Gyr ago. This region partially overlaps with the OCCS (off-centered compact source) identified by \citet{Johnston2024}, previously interpreted as a remnant of an external accretion.

In contrast, a control region along the bar minor axis (marked by a red circle in Figure~\ref{fig:malin1_rgb_fuv}) shows a typical old, super-solar, $\alpha$-poor stellar population consistent with a passively evolved disk.

Taken together, these results show remarkable chemical homogeneity among the C1 clump, SE arc, NW clump, and C0 region: all host metal-poor, $\alpha$-enhanced populations distinct from the surrounding disk. Despite their different current positions and velocities, their shared chemical properties suggest a common origin, possibly a disrupted dwarf galaxy group whose members are now at different stages of orbital decay or infall. 

The accreted dwarf is likely to have followed a high-inclination orbit. This configuration allows the infalling gas to remain relatively pristine and minimally contaminated by the pre-existing, metal-rich ($\sim$~solar) material in the Malin 1 disk. In contrast, a dwarf accreted along the disk plane would experience significant dynamical friction, causing it to spiral inward over several Gyrs or many rotation timescales. During this process, its gas would have ample time to mix with the enriched disk material, erasing the observed low-metallicity signature. Therefore, a recent accretion event along a high-inclination orbit offers the most plausible explanation for the chemical properties of the C1 clump.

\begin{figure*}[ht!]
\centering
\subfigure{\includegraphics[width=0.47\textwidth]{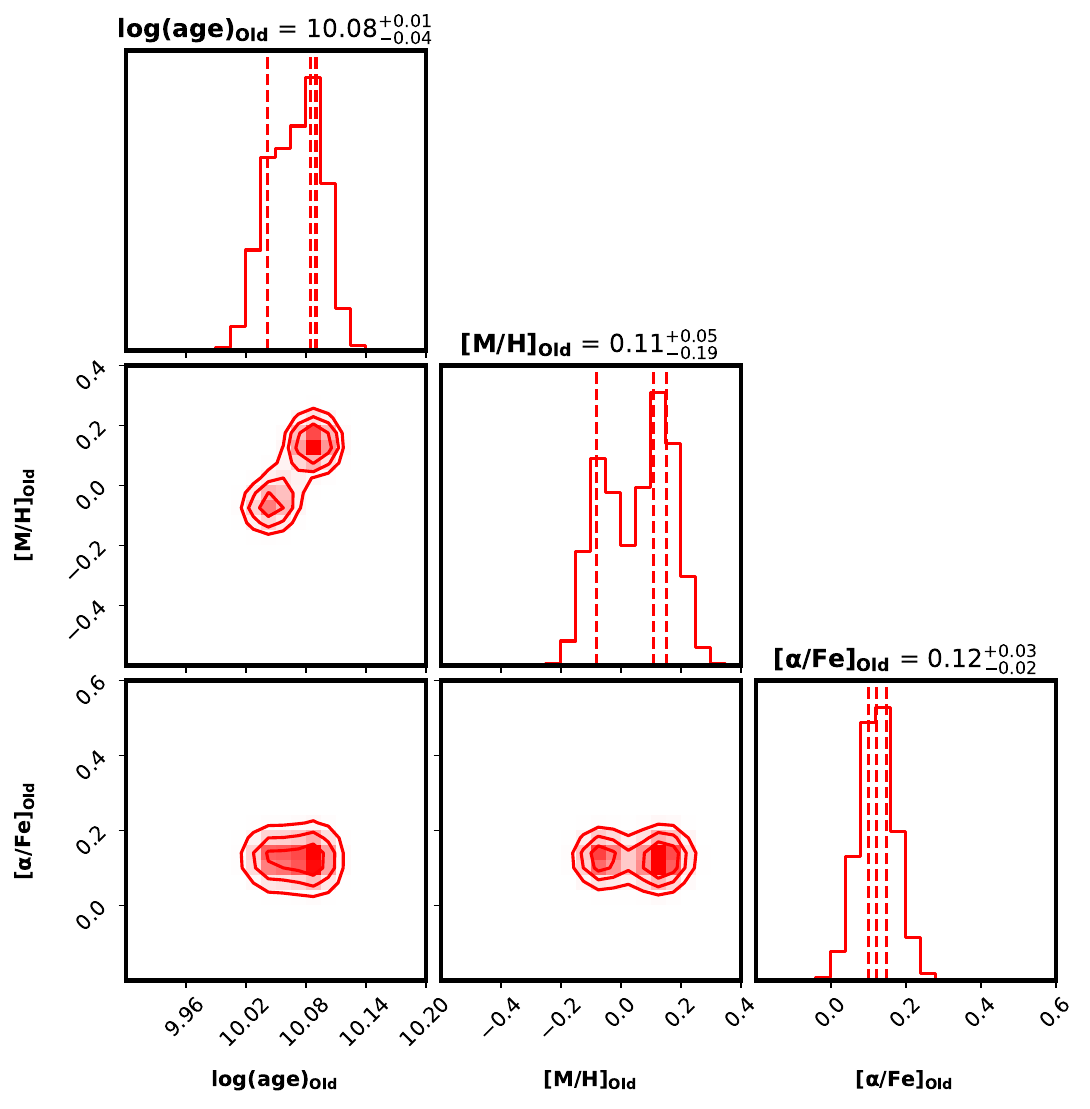}}
\subfigure{\includegraphics[width=0.47\textwidth]{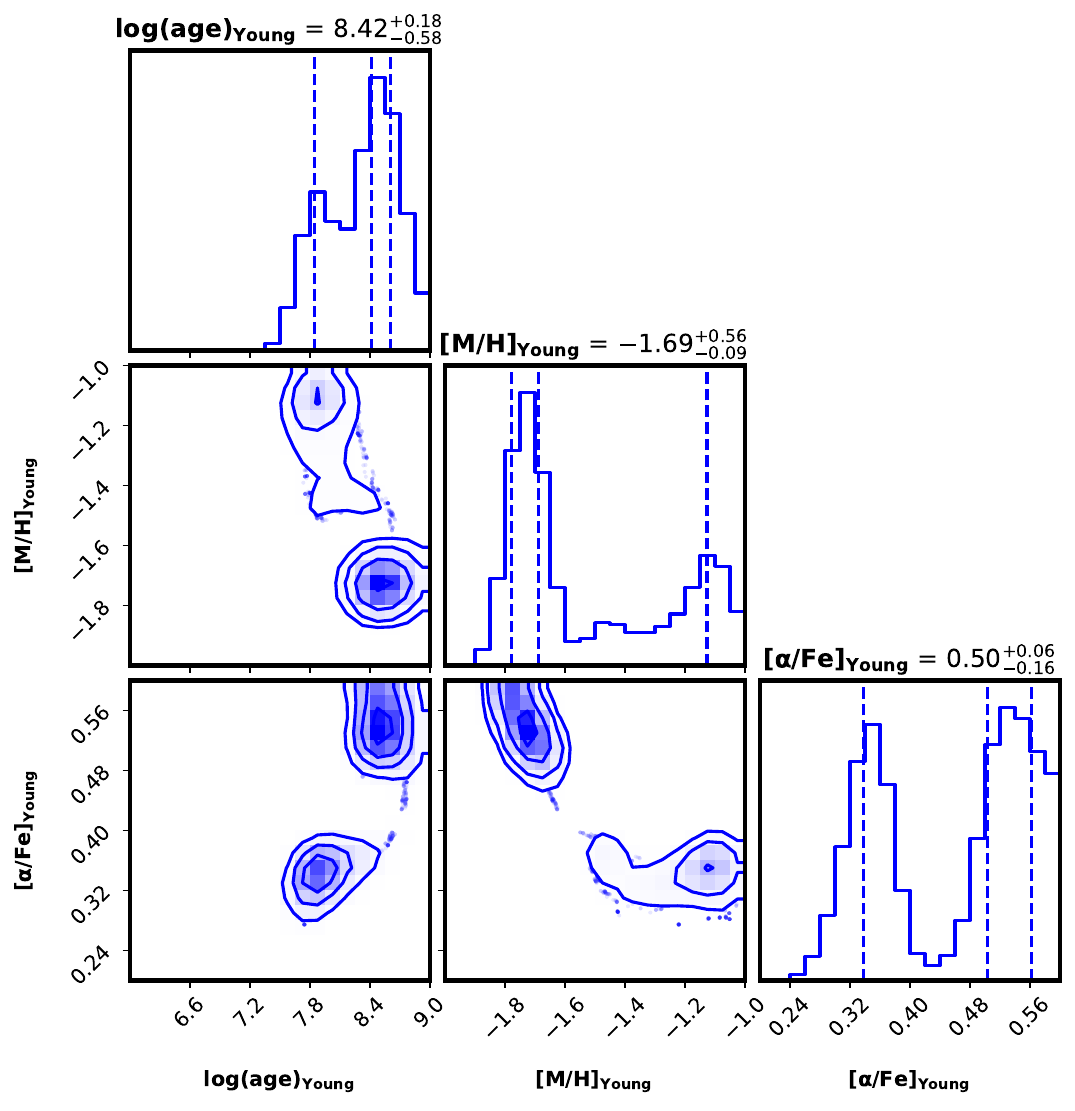}} \\
\caption{plot showing the corner plots of the Monte Carlo simulation of age, metallicity [M/H], and alpha enrichment $[\rm \alpha/Fe]$ resulting from 6240 realisations of the clump spectra. The red shows the parameter distribution of old stellar populations ($\rm age > 1Gyr$) and the blue one represents the younger population.
\label{fig:malin1_mcmc}}
\end{figure*}

\subsection{Monte Carlo simulations for C1 clump} 
\label{subsec:mc_analysis}

To assess the reliability of the pPXF stellar population fits, we performed a Monte Carlo (MC) analysis to quantify both statistical uncertainties and systematic effects related to the choice of regularization and multiplicative polynomial. By refitting the spectra under different combinations of these parameters, we evaluated how sensitive the derived stellar population properties are to the adopted fitting configuration.

We performed the Monte Carlo analysis on the C1 clump, as it exhibits the most distinct properties, including strong FUV emission and a significant velocity offset compared to the rotation of the Malin 1, consistent with a dwarf satellite accreting onto the central region.

After fitting the C1 clump spectra using pPXF, we performed a Monte Carlo simulation following the methodology of \citet{Pinna2019} (see Fig.~\ref{fig:malin1_mcmc}). First, we fit the observed spectrum without regularization or multiplicative polynomials. The residuals from this initial fit were modeled as a Gaussian with zero mean, and the resulting parameters were used to generate perturbed realizations of the original spectrum. We explored a grid of fitting configurations using six multiplicative polynomial degrees ($5 – 10$) and thirteen regularization values ($10^{-2}$ to 6). For each configuration, we produced 80 realizations, yielding a total of 6240 pPXF fits. From these, we computed the weighted average values for mass-weighted stellar age, metallicity ([M/H]), and alpha-enhancement ($[\alpha/\mathrm{Fe}]$). Figure~\ref{fig:malin1_mcmc} presents the posterior distributions in a corner plot, separating contributions from older ($>$1 Gyr; red) and younger ($\leq$1 Gyr; blue) stellar populations.

\noindent The overall population is dominated by old stars, with mass-weighted values: $\log[\text{age (Gyr)}] = 10.08^{+0.01}_{-0.04}$; $[\mathrm{M}/\mathrm{H}] = 0.10^{+0.04}_{-0.19}$;
$[\alpha/\mathrm{Fe}] = 0.12^{+0.02}_{-0.02}$.

\noindent In contrast, the young component shows:
$\log[\text{age (Gyr)}] = 8.42^{+0.18}_{-0.58}$;
$[\mathrm{M}/\mathrm{H}] = -1.69^{+0.56}_{-0.09}$;
$[\alpha/\mathrm{Fe}] = 0.50^{+0.06}_{-0.16}$, indicating a chemically distinct and externally originated stellar population. The overall distribution, not shown here, follows the older stellar population, as it is the dominant stellar population.

This clear separation between an old, metal-rich, alpha-poor population and a young, metal-poor, $\alpha$-enhanced population supports a two-phase assembly for the clump. The older stars are consistent with Malin~1’s central disk, while the younger stars — with their low metallicity ($[\mathrm{M}/\mathrm{H}] \sim -1.7$) and high $[\alpha/\mathrm{Fe}] \sim 0.5$, are inconsistent with internal gas recycling. Instead, they suggest enrichment by core-collapse supernovae in a chemically unevolved environment, consistent with a gas-rich dwarf progenitor. These findings strengthen the interpretation that this clump is the remnant of a recently accreted satellite, triggering localized star formation and contributing to ongoing inner disk growth in a system long thought to be quiescent.

The pPXF software performs full spectral fitting using a least-squares minimization approach to derive stellar kinematics and populations \citep{Cappellari2017}. However, spectral fitting can yield multiple solutions with comparably small $\chi^{2}$ values, particularly when age and metallicity exhibit strong degeneracy. In such cases, a fit with a slightly lower $\chi^{2}$ does not necessarily represent a more reliable or physically meaningful solution. During our Monte Carlo (MC) simulations, we examined the variation of the $\chi^{2}$ parameter, which was $\chi^{2} = 5132.2$ (dof = 2727) at zero regularization. We then adopted the criterion $\rm \chi^{2} = \chi^{2}_{0} + \sqrt{2\times dof}$, following \citet{Scott2021}, which was satisfied at a regularization value of $\rm regul = 3$ and a multiplicative polynomial degree of 8. These values lie within the explored grid parameters in our MC simulations. Hence, the application of regularization in pPXF effectively mitigates the degeneracy by favoring smoother and more physically consistent star formation histories.

\section{Conclusions} 
\label{sec:conclusion}
We have investigated the inner high surface brightness (HSB) disk assembly of the GLSB galaxy Malin 1 using deep ultraviolet imaging from AstroSat/UVIT and optical integral field spectroscopy from VLT/MUSE. Our analysis reveals multiple chemically primitive, kinematically distinct features in the central few kiloparsecs, suggesting that Malin 1 is actively growing its inner disk through the ongoing accretion of low-mass satellites. Our main findings are:

\begin{enumerate}

\item {UV Imaging \& Young Stellar Complexes:}\\
\textit{AstroSat} F154W imaging reveals clumpy, asymmetric far-UV emission across the central $\sim$10 kpc of Malin 1. The most prominent complex (C1), located southeast of the center, is visible only in F154W (not N263M), indicating a young, massive stellar population. Other features, including the NW complex, SE arc, and central OCCS, also show FUV and H$\alpha$ emission, tracing localized star formation within the last $\sim$10–100 Myr.

\item {Kinematics \& Dynamical Decoupling:}\\
MUSE observations reveal a strong localized velocity offset of $\sim 150\ \mathrm{km\ s^{-1}}$ in ionized gas at C1, along with elevated 
velocity dispersion ($\sim250\ \mathrm{km\ s^{-1}}$), both absent in the stellar velocity field. The NW complex and SE arc also appear as 
distinct features in H$\alpha$ velocity channel maps, suggesting that these structures are kinematically decoupled from the surrounding 
rotating disk and may be on plunging, off-plane orbits.

\item Chemically Distinct Stellar Populations:\\
Full spectral fitting using the sMILES library shows that the FUV-bright star-forming complexes C1, and NW, SE arc, and OCCS — host young ($\lesssim$1 Gyr), low-metallicity ([M/H] $\approx$ –1.7 to –0.75), and $\alpha$-enhanced stellar populations. This contrasts sharply with the surrounding bulge and bar-dominated regions, which are metal-rich ([M/H] $\gtrsim$ 0) and [$\alpha$/Fe]-poor, consistent with older stellar populations enriched primarily by Type Ia supernovae.

\item Chemical Homogeneity across SFCs:\\
Despite their different spatial locations and kinematic signatures, these SFCs share remarkably similar stellar population properties—low metallicity, elevated [$\alpha$/Fe], and young ages—suggesting a common external origin. We propose that they represent fragments of a disrupted dwarf galaxy currently being assimilated into Malin 1’s central region.

\item Implications for Disk assembly:\\
The presence of chemically primitive, kinematically decoupled star-forming structures embedded within an evolved SB0-type inner disk indicates that Malin 1 is actively assembling its central region through minor accretion events. These events deliver metal-poor, $\alpha$-enriched gas that fuels localized star formation, distinct from the host galaxy’s metal-rich, evolved stellar populations. This challenges the conventional view of GLSB galaxies as passively evolving systems and highlights that even massive S0-like galaxies can continue to grow through the accretion of low-mass galaxies.

\end{enumerate}

 \section*{Acknowledgment}

We thank the referee for their valuable comments, which greatly helped improve the quality of this paper. We thank Gulab Chand Dewangan and Nishant Singh for their useful discussion on this project. We gratefully acknowledge the Astrosat UVIT instrument, whose high resolution and sensitivity enabled this unique science case. We also acknowledge the Mikulski Archive for Space Telescopes (MAST) at the Space Telescope Science Institute (STScI) for maintaining and providing public access to the HST imaging datasets. We thank the ESO Science Archive for facilitating open access to the MUSE data, which forms the core of our analysis. Additionally, we acknowledge the IUCAA high-performance computing facility, Pegasus, for providing the computational resources essential for this work.

\appendix

\section{UVIT observations and analysis} 
\label{subsec:uvit_obs}
The Level 1 data in F154W and N263M bands were reduced using the official L2 pipeline. The pipeline was run orbit-wise, and later, each orbit's data was combined using a shift and rotation algorithm to produce a science-ready image. Details of Malin 1 data reduction are presented in \cite{Saha2021}. The photometric calibration is performed with a white dwarf star Hz4; the photometric zero-points are 17.78 and 18.18 for F154W and N263M, respectively \citep{Tandonetal2017a}. Once photometric calibration and astrometric correction are successfully applied, we extract an image of size 600" x 600" and run SExtractor~\citep{BertinArnouts1996} on it and extract the $3\sigma$ sources. After removing the $3\sigma$ sources, we place a large number of random apertures of $7\times7$ pixel boxes (each pixel = 0.417$^{\prime\prime}$), avoiding the location of the extracted sources. We fit a Gaussian to the resulting histogram (nearly symmetric) of fluxes in those boxes, and the mean of the Gaussian is considered the background. For F154W, the background is $B_{f}=9.5 \times 10^{-6}$~ct~s$^{-1}$~pix$^{-1}$ with a $\sigma_{f}=3.94 \times 10^{-5}$~ct~s$^{-1}$~pix$^{-1}$. In N263M filter, the sky background is $B_{n}=7.24 \times 10^{-5}$~ct~s$^{-1}$~pix$^{-1}$ with a $\sigma_{n}=6.48 \times 10^{-5}$~ct~s$^{-1}$~pix$^{-1}$ \citep{Saha2021}.
        
\cite{Saha2021} derived a UV star formation rate of $SFR_{FUV} = 0.094\pm0.03~M_{\odot}yr^{-1}$ within the central $2.6^{\prime\prime}$ region of Malin 1, adopting the calibration of \cite{Kennicutt98} and correcting for foreground extinction using \cite{Schlegel98}. Such a low value suggests that the galaxy's central region implies recent star formation, indicated by FUV and $H\alpha$ emission. Interestingly, the presence of offset FUV emission at the SE end of the bar points to localized activity and motivates a detailed understanding of the star formation history in the central region of Malin 1. Together, these findings raise the question of whether Malin 1 has experienced episodic rejuvenation or external perturbations that could have induced such recent star formation. Note that the inner HSB disk of Malin 1 has a radius of $9^{\prime\prime}$, having a magnitude of $20.79\pm0.04$ in FUV corresponding to $SFR_{FUV} = 0.416\pm0.017\ M_{\odot}yr^{-1}$.

\section{Archival imaging data from HST and CFHT} 
\label{subsec:hst_cfht_imaging}

We utilize the HST image of Malin 1 in the F814W/WFPC2 (PI: Chris Impey, \dataset[doi:http://dx.doi.org/10.17909/w9sj-xq50]http://dx.doi.org/10.17909/w9sj-xq50), via the Hubble Legacy Archive (HLA), with an exposure time of t$_{exp}$ = 2200 sec. CFHT images are obtained from Next Generation Virgo Cluster Survey (NGVS) \citep{Ferrarese2012} in the u, g, i, and z bands of Megacam, with t$_{exp}$ = 6402, 3170, 2055, and 3850 sec, respectively.

\section{MUSE Data reduction} 
\label{subsec:muse_data_reduction}
We obtained Multi Unit Spectroscopic Explorer (MUSE) integral field unit (IFU) data for Malin 1 from the ESO Science Archive (Program ID 105.20GH.001, PI: Galaz Gaspar). The observations consist of four on-source exposures, each 1160 s, along with two additional sets of offset sky frames. They were carried out in adaptive optics (AO)-assisted Wide Field Mode (WFM) with the extended wavelength setting, providing a $1^{\prime} \times 1^{\prime}$ field of view and a spatial sampling of $0.2^{\prime\prime}\mathrm{pix^{-1}}$. The MUSE wavelength coverage spans 4800–9300 \AA\ with a spectral sampling of 1.25 \text{\AA}$\mathrm{pix^{-1}}$. The total on-source exposure time is $t_{\mathrm{exp}} = 4640$ s, with an effective exposure of 4336 s after accounting for overheads. The data reach an AB magnitude of 25 at 5$\sigma$ sensitivity, with an average spectral resolution of $R \sim 3000$.

\begin{figure*}[ht!]
\centering
\includegraphics[width = 7in]{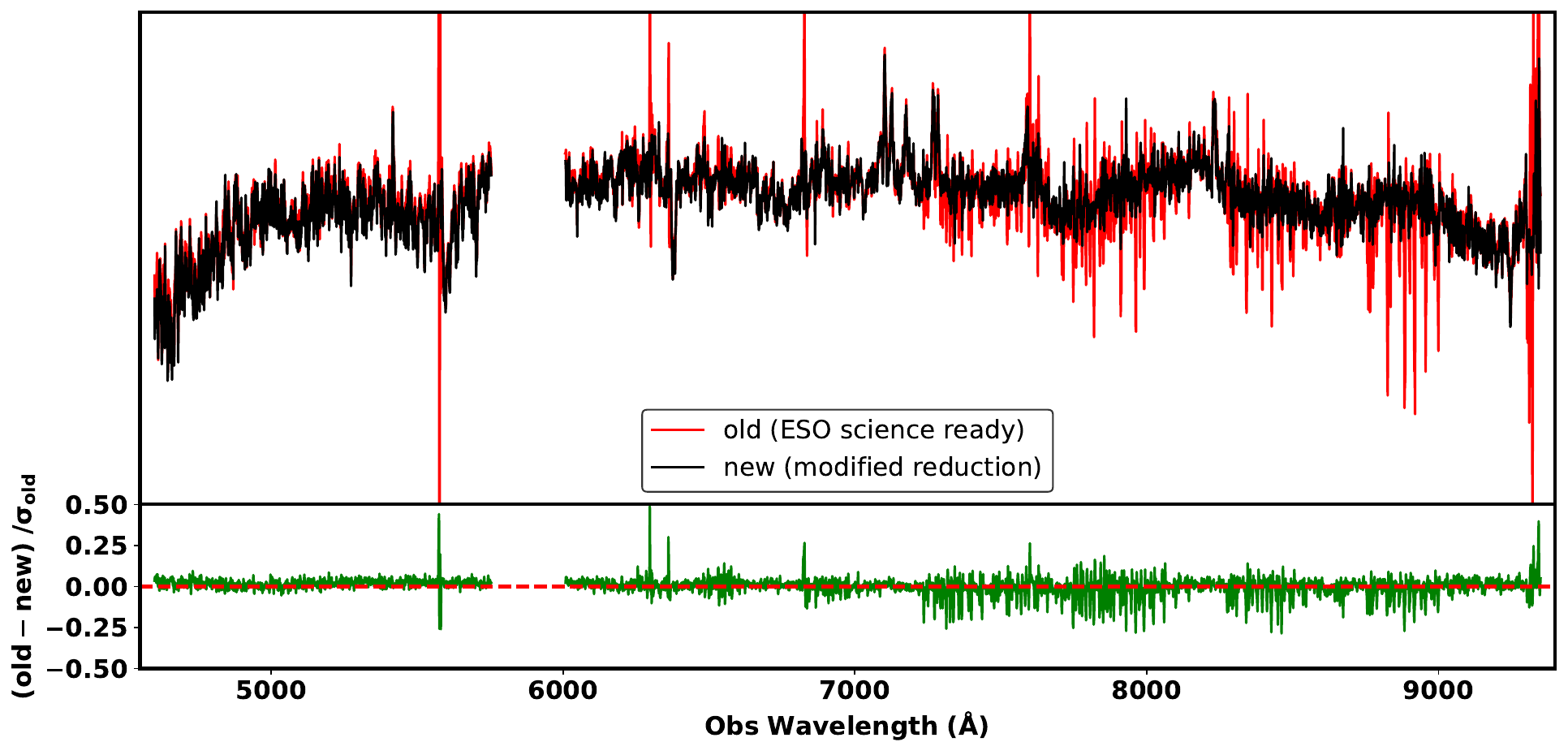}
\caption{Spectra of the FUV clumps are the same as what is modeled in the Figure \ref{fig:malin1_clump_c1_ppxf_fit}, here we show the spectra of the same clump from the science-ready data cube (red) provided through the ESO science portal, marked as old, and the spectra after reduction (black) as explained in the sec. \ref{subsec:muse_data_reduction}, marked as new. The bottom panel shows the residual of the old-new spectra divided by the $1\sigma$ error on the old spectra.
\label{fig:malin1_fuv_clump_old_new_comparison}}
\end{figure*}

We reduce the Malin 1 datacube with the help of the standard ESO MUSE pipeline \citep{Weilbacher2020} using the ESO Recipe Execution (EsoRex) Tool. We first prepare all the master files using the raw bias, dark, and flat files. Bias and dark frames are used to remove the CCD bias and the dark current in the detector. We then use the illumination frames closer in time to the science frames and the raw arc lamp file for the wavelength calibration. Standard star file provided with the raw data for flux calibration. We reduce and stack all the science files, taking into account the Raman emission lines. We don't use the offset sky observations or self-calibration due to known wavelength calibration issues. For sky subtraction, we apply the Zurich Atmosphere Purge (ZAP) and use the reduced science data itself with five iterations of ZAP to remove the sky \citep{Soto2016}. This has an advantage as it doesn't require the wavelength calibration to be accurate, and variations in the sky during offset sky observations. This procedure removes the sky lines and fringing in the redder wavelength part of the MUSE spectra, compared to using the MUSE standard sky subtraction (see Figure~\ref{fig:malin1_fuv_clump_old_new_comparison}). We correct the astrometry of the MUSE data using the HST/WFPC2/F814W image as a reference, as there was a systematic mismatch in the MUSE astrometry compared to HST archival imaging data. For astrometric correction, we use the same procedure and IDL program as used with UVIT data \citep{Saha2021}.

\begin{figure}[ht!]
\centering
\subfigure{\includegraphics[width = 2.12in]{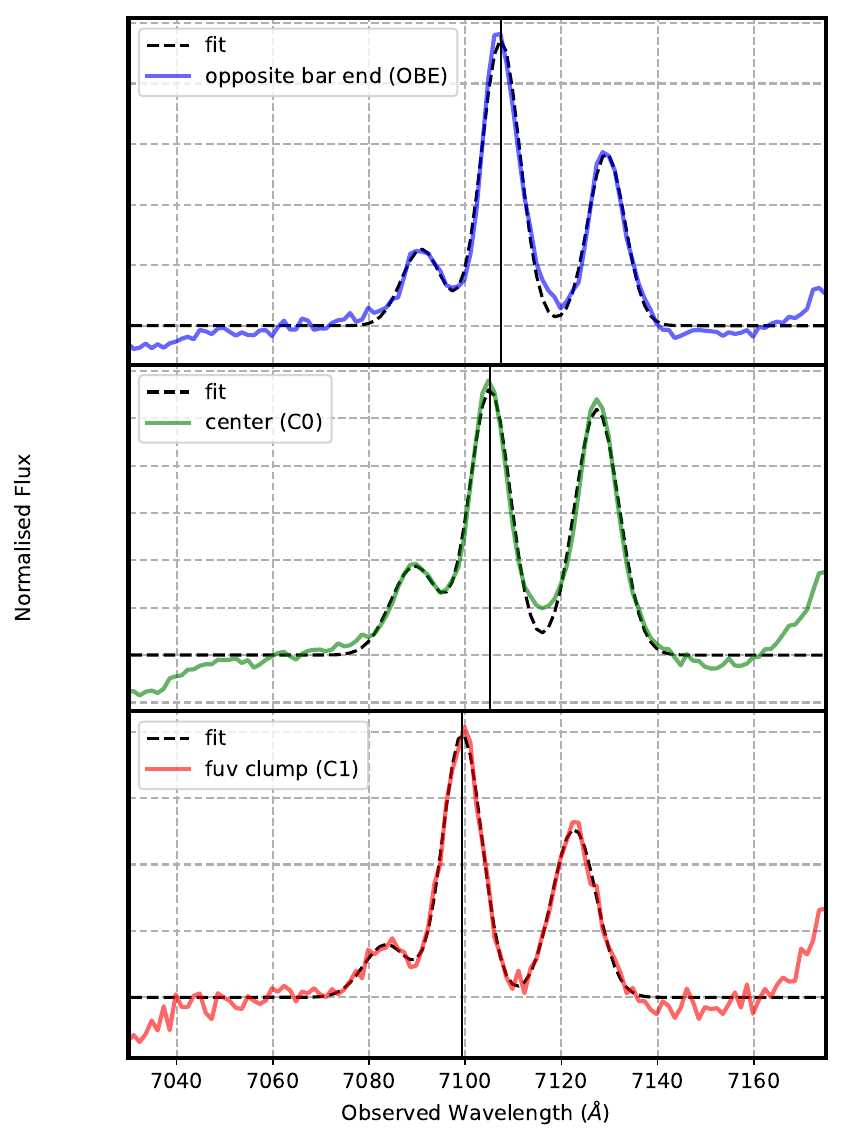}}
\subfigure{\includegraphics[width = 4in]{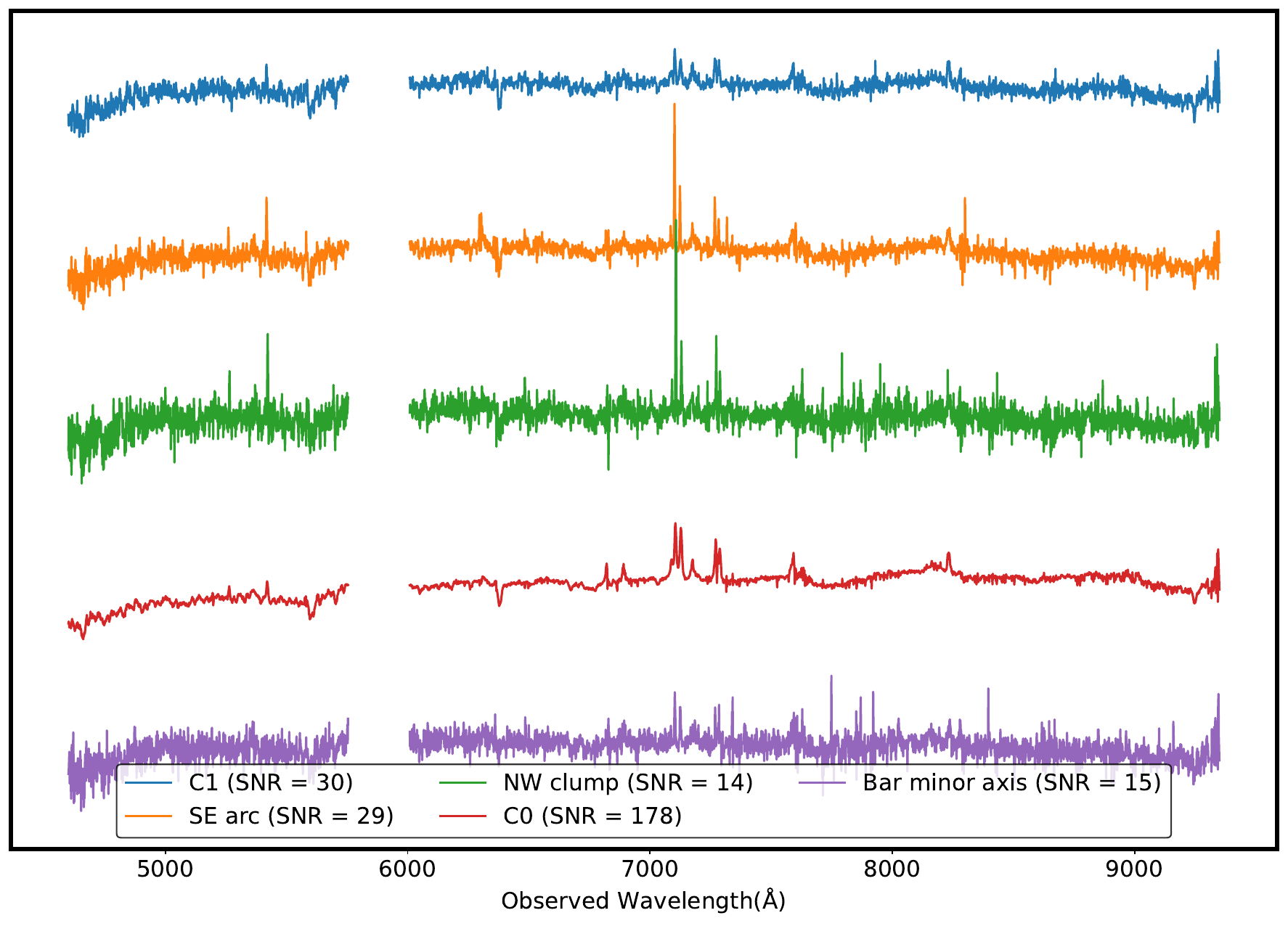}}
\caption{\textbf{Left:} shows the $\rm H\alpha$ emission from the positions indicated in Figure~\ref{fig:malin1_rgb_fuv}, marked by solid white circles. From this extraction, we measure an enhanced velocity of $\rm \sim 150\ kms^ {-1}$. \textbf{Right: }Extracted spectra for each feature. The spectra are normalized by median and offset by a constant for better visualization. 
\label{fig:malin1_extracted_spectra}}
\end{figure}

\subsection{MUSE analysis: Modeling spectra with pPXF } 
\label{subsec:ppxf_modeling}

We perform Voronoi binning \citep{Cappellari2003} to bin the science-ready data cube spatially to a target SNR=15 within the cube cutout covering the central HSB region. We bin the data cube based on the aggregated SNR of the $\rm H\alpha$ emission.

\begin{figure*}[ht!]
\centering
\includegraphics[width = 7in]{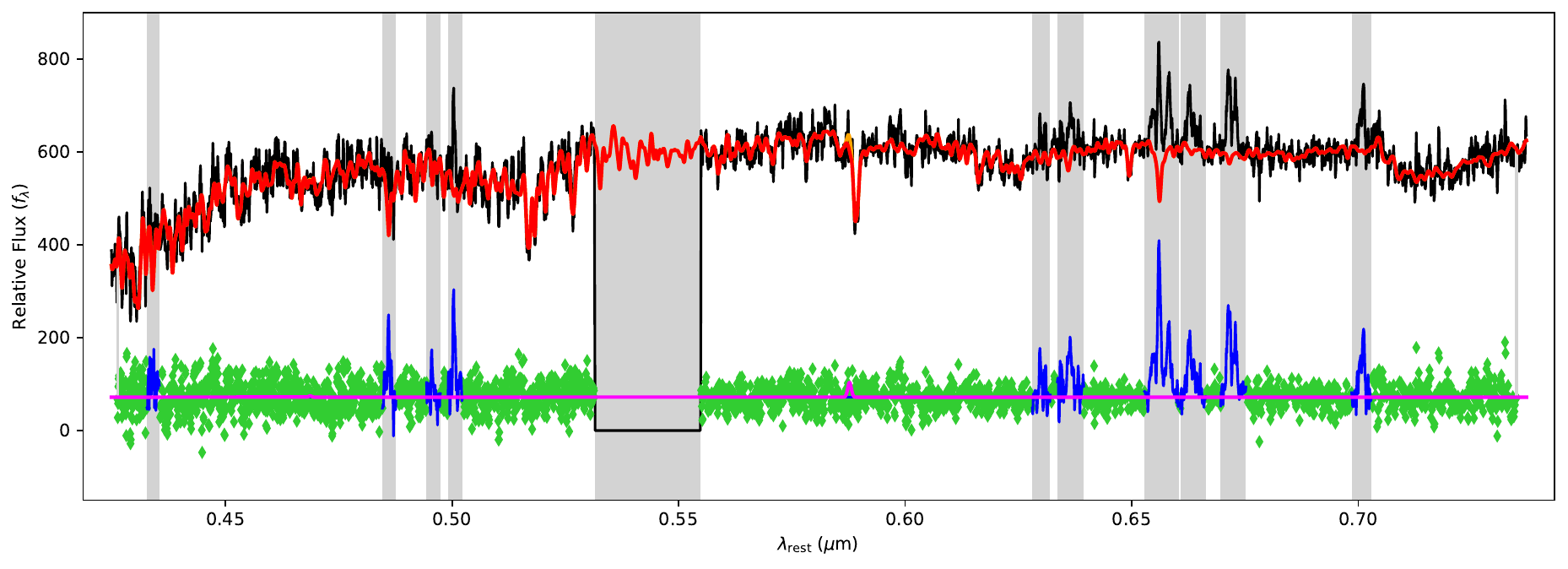}
\caption{pPXF fit to the FUV clump spectra shown in black, the stellar population model using sMILES is shown with red color. The vertical gray bands mark the region of potential emission lines and are masked while fitting; the broad masked region around the 0.55$\mu m$ is removed spectra using a notch filter to mask out the broad emission caused by sodium lasers used for AO. The green dots show the residual, while the blue color shows the masked spectra.
\label{fig:malin1_clump_c1_ppxf_fit}}
\end{figure*}

We utilize the semi-empirical MILES (sMILES, \cite{Knowles2023}) SSP library, with wavelength coverage of 3540.5 - 7409.6 \text{\AA}, to model the extracted spectra (see Figure~\ref{fig:malin1_extracted_spectra} for spectra, Figure~\ref{fig:malin1_clump_c1_ppxf_fit} for model fit) with the help of pPXF \citep{Cappellari2004, Cappellari2017}. This stellar population library contains a grid of semi-empirical, single-age, single-metallicity stellar population (SSP) model spectra at five different [$\alpha$/Fe] bins, with [$\alpha$/Fe] ranging from -0.2 to 0.6; with 53 age bins ranging from 0.03 - 14 Gyr, 10 metallicity bins with [M/H] ranging from -1.79 to 0.26. Additionally, these models are computed for five different IMFs, but for this study, we only use the unimodal Salpeter IMF \citep{Salpeter1955} with a logarithmic slope of 1.3.

Using the pPXF spectral model fitting, we derive the stellar and gas kinematic maps of the central region of the galaxy. The pPXF Python module extracts the stellar/gas kinematics and stellar population from absorption/emission line spectra of galaxies via a maximum penalized likelihood approach. We use the EMILES stellar population library \citep{Vazdekis2016} to derive kinematic maps, as its model spectra span 1680–49999 \text{\AA} in wavelength. For the given redshift and MUSE spectral range, this coverage allows us to include all wavelengths present in our data, ensuring that all absorption and emission lines are available for spectral fitting. This comprehensive coverage allows for more reliable kinematic measurements, particularly of stellar velocities.

\begin{figure}[ht!]
\centering
\subfigure{\includegraphics[width = 3.5in]{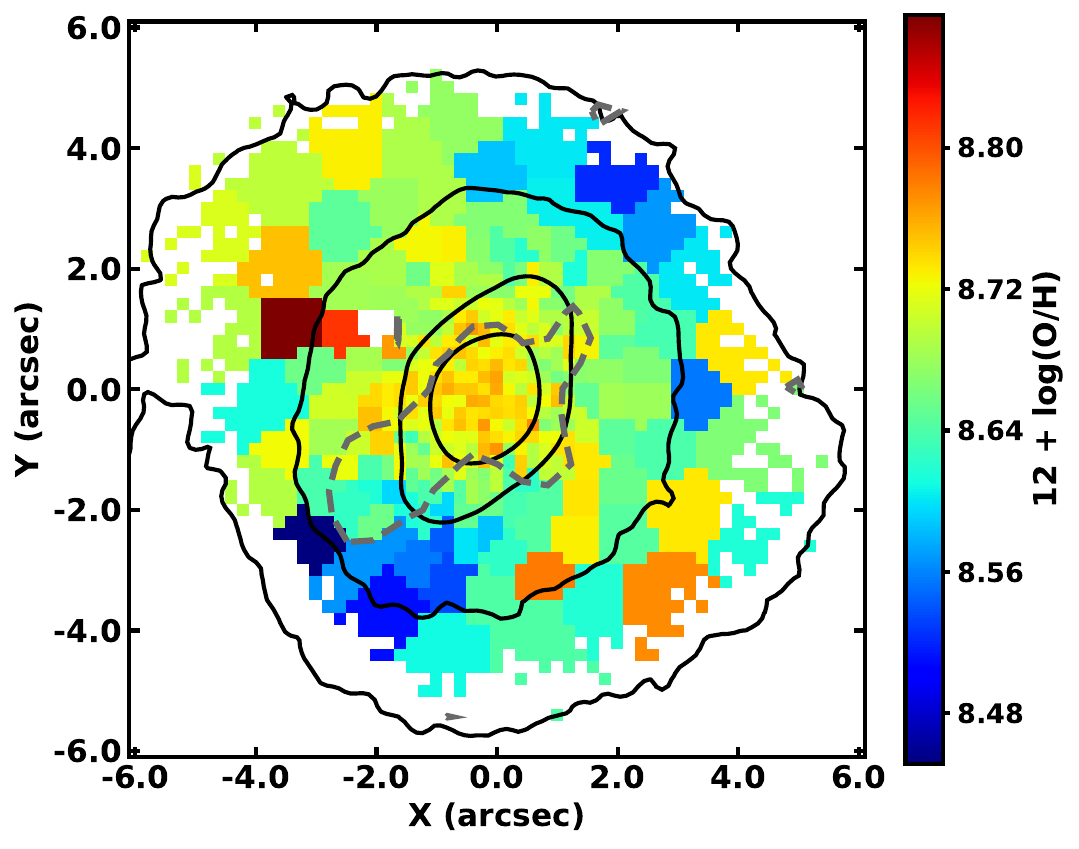}}
\subfigure{\includegraphics[width = 3.5in]{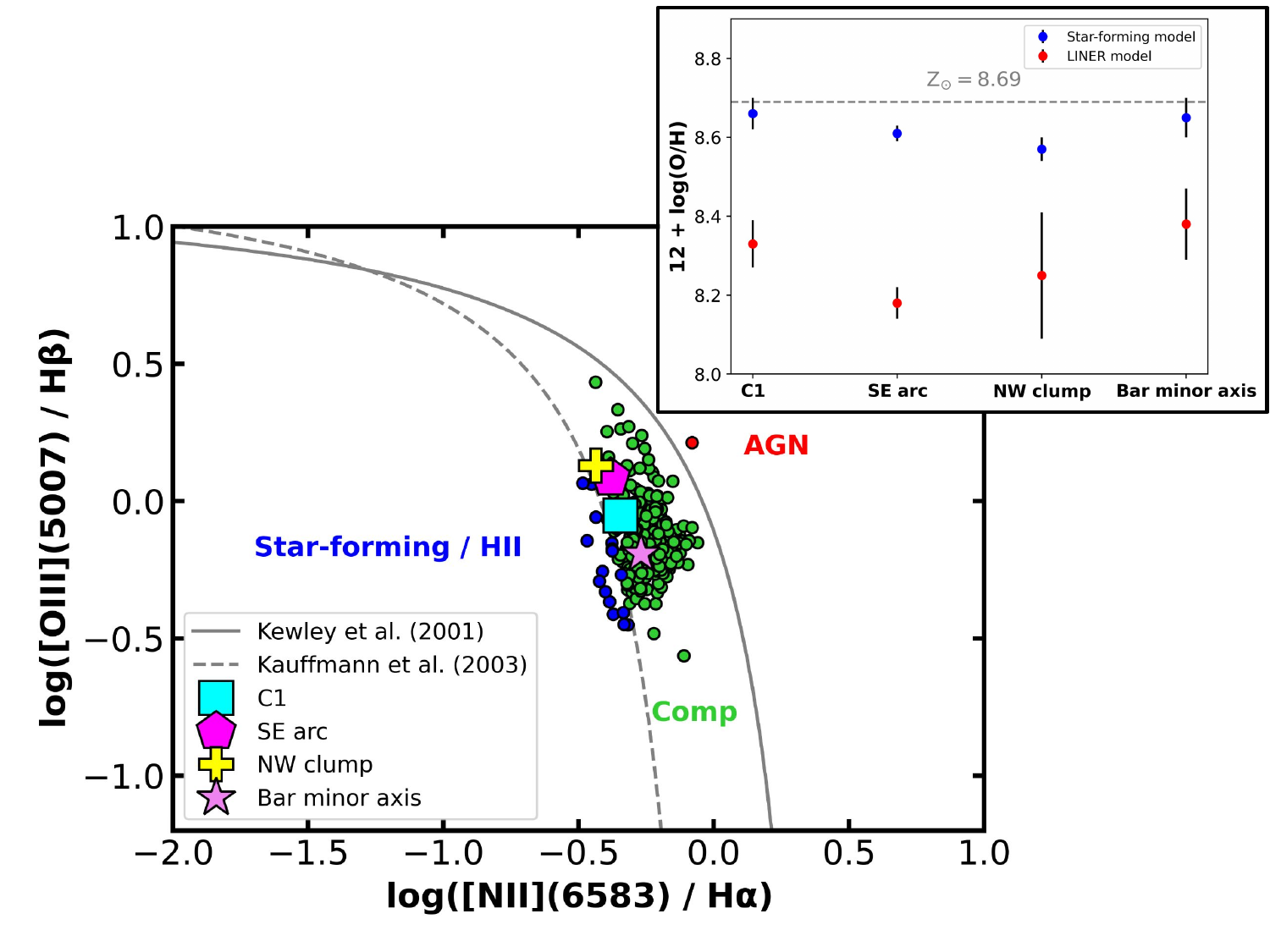}}
\caption{\textbf{Left: }Resolved gas-phase metallicity map of the central region of Malin 1 derived using the O3N2 calibration. \textbf{Right: }Spatially resolved BPT diagram for the bins within the central region, with the four different regions shown using distinct symbols as indicated in the legend. The inset (top right) shows the gas-phase (O/H) abundances derived from photoionization models for the clumps analyzed in this study. Blue points correspond to star-forming regions, while red points indicate regions ionized by LINER-like sources. The dashed line marks the solar metallicity.}
\label{fig:malin1_o3n2_metallicity_map}
\end{figure}

\clearpage
\bibliography{malin1_ref}{}
\bibliographystyle{aasjournal}

\end{document}